\documentclass{article}
\usepackage{spconf,amsmath,graphicx,multirow,amsfonts,amssymb}
\newcommand{\etal}{{\emph{et al.}}}
\newcommand{\ie}{{\emph{i.e.}}}
\newcommand{\eg}{{\emph{e.g.}}}

\title{Towards simultaneous segmentation of liver tumors and intrahepatic vessels via cross-attention mechanism}
\name{Haopeng Kuang\textsuperscript{1} \qquad Dingkang Yang\textsuperscript{1} \qquad Shunli Wang\textsuperscript{1} \qquad Xiaoying Wang\textsuperscript{2} \qquad Lihua Zhang \textsuperscript{1,3*}
\thanks{*Corresponding author. This project is funded by the National Natural Science Foundation of China (82090052,82090054)}}
\address{\textsuperscript{1}Academy for Engineering \& Technology, Fudan University, Shanghai, China \\\textsuperscript{2}Liver Surgery Department, Zhongshan Hospital, Fudan University, Shanghai, China
\\\textsuperscript{3}Engineering Research Center of AI \& Robotics, Ministry of Education, Shanghai, China}

\begin{document}

\maketitle
\begin{abstract}
Accurate visualization of liver tumors and their surrounding blood vessels is essential for noninvasive diagnosis and prognosis prediction of tumors. In medical image segmentation, there is still a lack of in-depth research on the simultaneous segmentation of liver tumors and peritumoral blood vessels. To this end, we collect the first liver tumor, and vessel segmentation benchmark datasets containing 52 portal vein phase computed tomography images with liver, liver tumor, and vessel annotations. In this case, we propose a 3D U-shaped Cross-Attention Network (UCA-Net) that utilizes a tailored cross-attention mechanism instead of the traditional skip connection to effectively model the encoder and decoder feature. Specifically, the UCA-Net uses a channel-wise cross-attention module to reduce the semantic gap between encoder and decoder and a slice-wise cross-attention module to enhance the contextual semantic learning ability among distinct slices. Experimental results show that the proposed UCA-Net can accurately segment 3D medical images and achieve state-of-the-art performance on the liver tumor and intrahepatic vessel segmentation task.
\end{abstract}

\section{Introduction}
\label{sec:intro}

Benefiting from the development of deep learning techniques \cite{yang2022contextual, yang2022disentangled, yang2022learning, yang2023target}, the development of hepatectomy has entered a new stage of high precision.
The relationship between liver tumors and intrahepatic blood vessels directly determines whether to use surgical resection and how to formulate high-quality surgical plans.
Most liver tumor and vessel segmentation research are based on the Liver Tumor Segmentation (LiTS) challenge \cite{LiTS2017}, the Medical Segmentation Decathlon (MSD) challenge \cite{MSD2022}, and the 3D-IRCADB dataset \cite{3DIRCADB2010}. These datasets only consider the independent segmentation task of liver tumors or blood vessels, and the segmentation accuracy of liver vasculature is not high enough. The 3D-IRCADB dataset \cite{3DIRCADB2010} released in 2010 with both liver tumor and intrahepatic vascular labels has only 20 cases. In this case, the inadequate available data significantly limits researchers' ability to improve existing models.

Most existing studies on the segmentation of liver tumors \cite{E2Net2020,3DVA2019,LWHCN2019} and vessels \cite{kuang2022,2dresunet2018} are based on U-shaped encoder-decoder networks \cite{UNET2015, 3DAttentionUnet2021} (U-Net). Among various improvements to the U-Net, representative studies are the design of a two-stage framework \cite{E2Net2020}, the introduction of the attention module \cite{3DVA2019}, and the 2D and 3D hybrid convolution structure \cite{LWHCN2019}.
However, Wang~\etal~\cite{UCTransNet2022} proves the effectiveness of the U-Net framework on multiple datasets and confirms that the skip connection is not all beneficial for semantic segmentation. This is because the encoder and decoder phases' feature sets are incompatible, and the simple concatenation operation will make some skip connections harm the segmentation performance. Therefore, Wang~\etal~ \cite{UCTransNet2022} propose a scheme to fuse a Transformer \cite{selfAtt2022} and 2D U-Net to remove the limitation of skip connection. Considering that the cross-attention \cite{CAT2022} allows information interaction between different feature maps, this paper re-thinks the skip connection of 3D U-Net \cite{3Dunet2016}. More concretely, our insight is to introduce the cross-attention module to replace the skip concatenation while preserving the efficient encoder-decoder structure of U-Net, to fuse the encoder features better and reduce the semantic gap between the encoder and decoder.

\begin{figure}[t]
  \centering
  \centerline{\includegraphics[width=9cm]{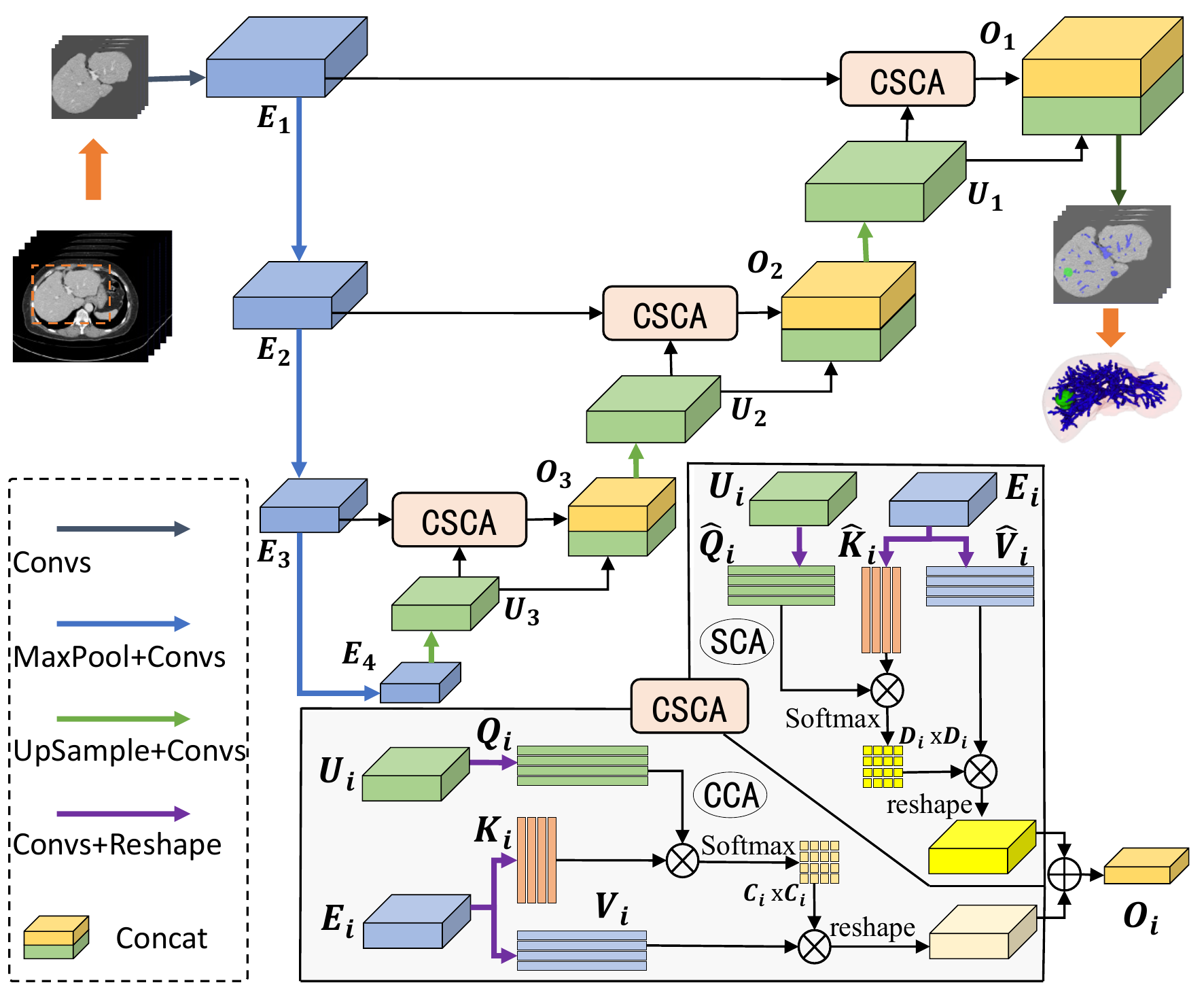}}
%  \vspace{2.0cm}
%   \centerline{}\medskip
\caption{Illustration of the proposed UCA-Net, where the CSCA module consists of two components: the CCA and SCA modules.}
\label{fig:CSCA}
\end{figure}

This paper proposes an elegant and efficient 3D U-shaped Cross-Attention Network (UCA-Net) to replace the skip connection using a cross-attention mechanism. The network architecture of UCA-Net is shown in Figure \ref{fig:CSCA}. Each time after up-sampling, the decoder will take advantage of the cross-attention to exchange information from the down-sampled features of the encoder at the corresponding stage. The primary function of the Channel-wise Cross-Attention (CCA) module is to fuse the encoder and decoder features and reduce the semantic gap between them. The Slice-wise Cross-Attention (SCA) module is designed to enhance further the learning ability of 3D context semantics among slices.
In addition, we collect the portal vein CT data of 52 patients with liver tumors and invite professional doctors to label the liver, intrahepatic tumors, and hepatic vessels. The collected data is named the Liver Tumor and Vessel Segmentation (LTVS) dataset \footnote{Data are available from the corresponding author for researchers.}, which can promote the research progress of liver tumor and vessel segmentation tasks.

\section{Material And Methodology}
\label{sec:MaterialAndMethods}

\subsection{Dataset}
\label{ssec:Dataset}
To better serve liver surgery planning, we collected LTVS datasets to advance the task of simultaneous segmentation of tumor and peritumoral vessels. We collected the portal vein phase abdominal CT image data of 52 patients with liver tumors from different devices between January 2020 and February 2021. The liver surgery department of Zhongshan Hospital, Fudan University, has approved this study. The thickness of these data is 1mm in 33 patients, 0.8mm in 14 patients, 0.7999mm in 4 patients, and 2mm in 1 patient. The spacings of these data range from 0.6190mm to 0.8223mm in the axial plane. In addition, these 52 CT volume data contain 16,052 slices of $512\times512$ pixel size, with each CT volume data containing between 157 and 501 slices.

\begin{figure}[t]
  \centering
  \centerline{\includegraphics[width=7.5cm]{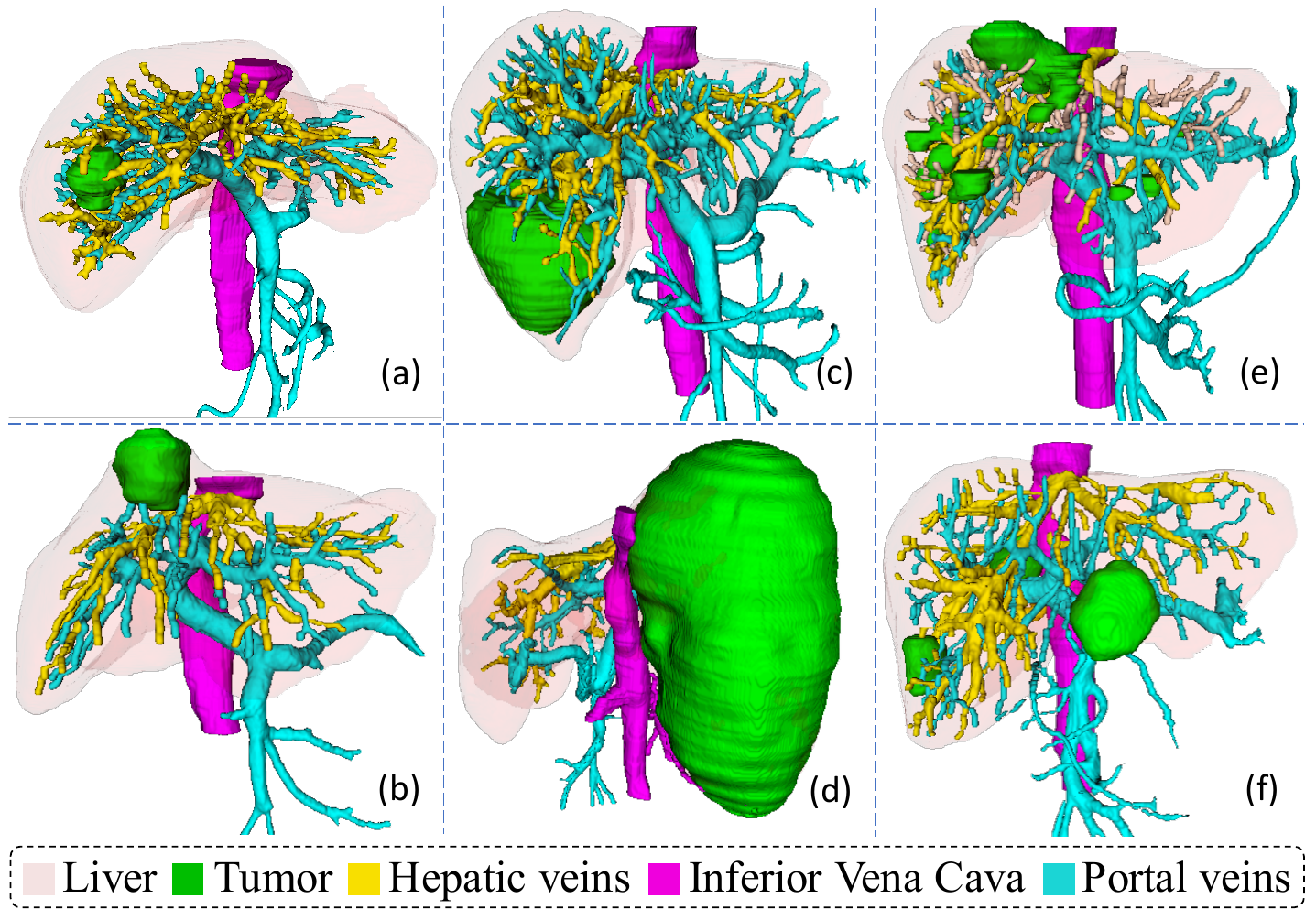}}
%  \vspace{2.0cm}
%   \centerline{}\medskip
\caption{Visual renderings of representative examples. We highlight single versus multiple tumors, small versus large tumors, and intrahepatic tumors versus tumors that breach the boundaries of the liver.}
\label{fig:LTVS}
\end{figure}

Experienced radiologists manually annotated our LTVS dataset with two iterations of revision. In addition to accurately labeling liver and liver tumors, radiologists made detailed labeling of hepatic vessels according to hepatic veins, inferior vena cava, and portal veins. Figure \ref{fig:LTVS} shows a visual rendering of some representative data.
Firstly, our dataset is more timely compared with previous datasets. The research based on our dataset can be more easily integrated with current clinical needs.
Secondly, our carefully hand-picked data is rich in diversity. The location, number, and size of tumors in our data are different, which is conducive to further evaluation and improvement of the segmentation model.
Finally, the annotation of the LTVS dataset is more refined. Our labeling of vessels is very accurate and includes the entire hepatic venous system, which can not only support the independent segmentation of the liver, vessels, and tumors but also promote the accurate segmentation of tumors and peritumoral vessels.

\subsection{Methodology}
\label{ssec:methodology}

\subsubsection{Data Preprocessing}
\label{sssec:Preprocessing}
To accurately segment liver tumors and peritumoral vessels, we focus on precise segmentation of liver tumors and intrahepatic small blood vessels. To this end, we first use the ground truth of the liver to remove the labels of vessels located outside the liver. The removed vessels are mainly the inferior vena cava and part of the portal veins with large diameters and easy segmentation.
Then, we perform threshold truncation and normalization of CT values for 3D volume data according to the standard process of CT image segmentation task, and the threshold range is set as [-160, 240] according to doctors' experience.

\subsubsection{Liver Segmentation}
\label{sssec:liver}
We uniformly sample the axial spacing of 3D data to 1mm and reduce the sampling from $512\times512$ to $320\times320$ for each slice. To reduce GPU requirements, we use patch overlap mechanism (overlap slice: 16) to extract $32\times320\times320$ patches from the data and feed them into the neural network based on 3D U-Net. In prediction, we use the same mechanism to assemble all the patch prediction results into the prediction results of the whole 3D volume data. It is important to note that we consider the tumor and intrahepatic blood vessels as a part of the liver when performing liver segmentation.

\begin{table}[t]
\small
\centering
\renewcommand{\arraystretch}{1.2}
\caption{Comparison results on the LTVS dataset. The bold font indicates that our method outperforms the others. ``$^\star$" denotes the method with a transformer structure, where the pre-trained parameters are used for the TransUnet.}
\label{Tab1}
\vspace{3pt}
\resizebox{1\linewidth}{!}{%
\begin{tabular}{ccccccc}
\hline
\multirow{2}{*}{Methods} & 
\multicolumn{1}{c}{Params} & \multicolumn{2}{c}{Liver Tumor} & \multicolumn{2}{c}{Intrahepatic Vessel} \\
\cline{3-6}
& (M) & Dice(\%) & IoU(\%)  &  Dice(\%) & IoU(\%)     \\ 
\hline\hline
\multicolumn{1}{c|}{2D Unet \cite{UNET2015}}       &  2.14 & 76.26 & 64.89 & 60.30 & 44.10 \\
\multicolumn{1}{c|}{2D ResUnet \cite{2dresunet2018}}    &  3.26 & 76.30 & 64.87 & 62.45 & 45.87 \\
\multicolumn{1}{c|}{2D UCTransNet \cite{UCTransNet2022}}    &  65.60 & 79.09 & 66.26 & 62.62 & 46.26 \\
\multicolumn{1}{c|}{2D TransUnet$^\star$ \cite{transunet2021}} & 105.48 & 77.99 & 64.19 & 62.39 & 45.63 \\
\hline
\multicolumn{1}{c|}{3D Unet \cite{3Dunet2016}}     &  6.42 & 80.36 & 68.96 & 66.00 & 49.10 \\
\multicolumn{1}{c|}{3D ResUnet \cite{3dresunet2019}} &  9.43 & 83.71 & 73.50 & 66.97 & 50.71 \\
\multicolumn{1}{c|}{3D Attention Unet \cite{3DAttentionUnet2021}}    & 16.50 & 83.01 & 73.20 & 66.15 & 50.33 \\
\multicolumn{1}{c|}{3D Vnet \cite{Vnet2016}}    & 45.60 & 83.60 & 72.65 & 65.53 & 49.98 \\
\multicolumn{1}{c|}{3D UX-Net$^\star$ \cite{uxnet2022}}  & 57.43 & 82.81 & 72.04 & 66.76 & 50.40 \\
\multicolumn{1}{c|}{\textbf{Ours}}        &  5.91 & \textbf{84.96} & \textbf{74.38} & \textbf{67.47} & \textbf{51.34} \\
\hline
\end{tabular}
}
%\vspace{-1pt}
\end{table}

\subsubsection{Liver Tumor $\&$ Intrahepatic Vessel Segmentation}
\label{sssec:Tumor}
We utilize the segmented liver region as the input of UCA-Net, a segmentation network with a size of $32\times320\times320$. Since each person's liver size differs, we still use the same resampling and patch overlap mechanism for the liver region as in Section \ref{sssec:liver}. The main improvement of the proposed UCA-Net is the introduction of the CSCA module consisting of CCA and SCA.

\noindent\textbf{Channel-wise Cross-attention Module (CCA).}
To better fuse the semantically inconsistent features between the encoder and decoder, we propose a channel-wise cross-attention module, which can guide the information filtering on the encoder features' channel and eliminate the decoder's ambiguity features caused by upsampling.

In terms of a mathematical representation, we take the feature map $\mathbf{E}_i$ of the $i$-th $(i= 1,2,3)$ level encoder and the upsampled feature map $\mathbf{U}_i$ of the $(i+1)$-th level decoder as the input of the CCA module. We first project the feature map $\mathbf{U}_i$ to space $\mathbf{Q}$ as queries and the feature map $\mathbf{E}_i$ to space $\mathbf{K}$ and $\mathbf{V}$ as keys and values:
\begin{equation}\label{CCA_1}
   \mathbf{Q}_i=\mathbf{W}_{\mathbf{Q}_i} \mathbf{U}_i,
   \mathbf{K}_i=(\mathbf{W}_{\mathbf{K}_i} \mathbf{E}_i)^T,
   \mathbf{V}_i=\mathbf{W}_{\mathbf{V}_i} \mathbf{E}_i,
\end{equation}
where $\mathbf{W}_{\mathbf{Q}_i}\in \mathbb{R}^{C_i\times L_i}, \mathbf{W}_{\mathbf{K}_i}\in \mathbb{R}^{C_i\times L_i}, \mathbf{W}_{\mathbf{V}_i}\in \mathbb{R}^{C_i\times L_i}$, are learnable parameters, $C_i$ is the channel dimension of the $i$-th layer, and $L_i$ is the length of the sequence. In our experiment, $L_i$ is calculated by the size of $\mathbf{E}_i$.
Through the cross-attention mechanism, we can get the similarity matrix $\mathbf{M}_i $ and multiply it with $\mathbf{V}_i$ to get the output $CCA_i$:
\begin{equation}\label{CCA_2}
   CCA_i=\mathbf{M}_i \mathbf{V}_i=\sigma(\mathbf{Q}_i \mathbf{K}_i)\mathbf{V}_i,
\end{equation}
where $\sigma \left(\cdot\right)$ is the softmax function.

\noindent\textbf{Slice-wise Cross-attention Module (SCA).}
To improve the learning ability of slice context semantics in 3D volume data, we propose a slice-wise cross-attention module, which enables the decoder to fuse the slice context semantics of filtered encoder features.

Like CCA, $\mathbf{E}_i$ and $\mathbf{U}_i$ are the input of the SCA module. The difference is that focusing attention on the slice context semantics representation. We compress the dimension of the channel to one when projecting features to the space $\mathbf{\hat{Q}},\mathbf{\hat{K}}$ and $\mathbf{\hat{V}}$:
\begin{equation}\label{SCA_1}
   \mathbf{\hat{Q}}_i=\mathbf{W}_{\mathbf{\hat{Q}}_i} \mathbf{U}_i,
   \mathbf{\hat{K}}_i=(\mathbf{W}_{\mathbf{\hat{K}}_i} \mathbf{E}_i)^T,
   \mathbf{\hat{V}}_i=\mathbf{W}_{\mathbf{\hat{V}}_i} \mathbf{E}_i,
\end{equation}
where $\mathbf{W}_{\mathbf{\hat{Q}}_i}\in \mathbb{R}^{D_i\times \hat{L}_i}, \mathbf{W}_{\mathbf{\hat{K}}_i}\in \mathbb{R}^{D_i\times \hat{L}_i}, \mathbf{W}_{\mathbf{\hat{V}}_i}\in \mathbb{R}^{D_i\times \hat{L}_i}$, are the weights of different inputs, $D_i$ is the depth of $\mathbf{E}_i$, and $\hat{L}_i$ is calculated by multiplying the width and height of $\mathbf{E}_i$.
Similarly, $SCA_i$ can be represented as:
\begin{equation}\label{SCA_2}
   SCA_i=\sigma(\mathbf{\hat{Q}}_i \mathbf{\hat{K}}_i)\mathbf{\hat{V}}_i.
\end{equation}
In particular, we merge data by the channel during projection mapping, where the channel dimension is 1. And the channel dimension is restored at the end of the SCA module.
Finally, we can get the output of the CSCA module
$\mathbf{O}_i=CCA_i \oplus SCA_i,$
where $\oplus$ denotes the element-wise addition operation.

\begin{table}[t]
\small
\centering
\renewcommand{\arraystretch}{1.2}
\caption{Ablation study results.``w/o" means without,``w/" means with, and the bold font shows the best results.}
\label{Tab2}
\vspace{3pt}
\resizebox{1\linewidth}{!}{%
\begin{tabular}{ccccccc}
\hline
\multirow{2}{*}{Methods} & \multicolumn{1}{c}{Params} & \multicolumn{2}{c}{Liver Tumor} & \multicolumn{2}{c}{Intrahepatic Vessel} \\
\cline{3-6}
& (M) & Dice(\%) & IoU(\%)  &  Dice(\%) & IoU(\%)     \\ 
\hline\hline
\multicolumn{1}{c|}{w/o CSCA }     &  6.42   & 80.36 & 68.96 & 66.00 & 49.10 \\
\multicolumn{1}{c|}{w/ SCA }       &  5.84 & 84.20 & 73.47 & 65.16 & 48.07 \\
\multicolumn{1}{c|}{w/ CCA }       &  5.90 & 82.73 & 71.90 & 67.40 & 51.26 \\
\multicolumn{1}{c|}{\textbf{Ours}} &  5.91 & \textbf{84.96} & \textbf{74.38} & \textbf{67.47} & \textbf{51.34} \\
\hline
\end{tabular}%
}
\end{table}

\section{EXPERIMENTS AND RESULTS}
\label{sec:EXPERIMENTS}

\begin{figure*}[t]
  \centering
  \centerline{\includegraphics[width=18cm]{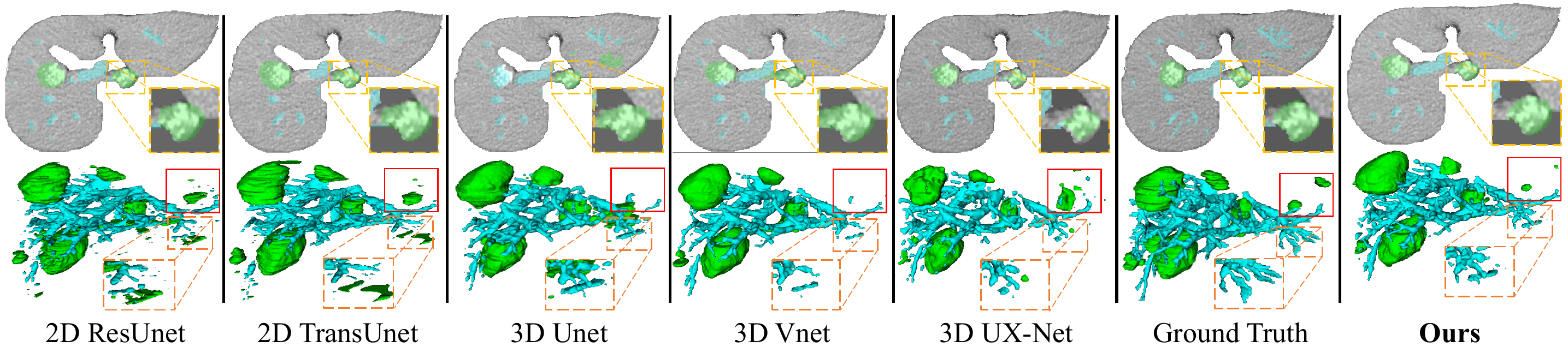}}
%  \vspace{2.0cm}
%   \centerline{}\medskip
\caption{Visualization of segmentation results of liver tumors and vessels. The gold dotted box shows the segmentation results in the 2D slices. Small tumors and terminal vessels are shown in the 3D view in solid red and dashed orange boxes, respectively.}
\label{fig:PRED}
\end{figure*}

\subsection{Implementation Details}
\label{ssec:Implementation}
We use the identical setup and dice loss function \cite{kuang2022} to train all baselines. Specifically, we implement our proposed model using PyTorch on two NVIDIA V100 GPUs with 32 GB of memory. For data augmentation, we only use random rotation operations to avoid overfitting. The LTVS dataset is divided into training, validation, and testing sets in a 7:1:2 ratio (\ie, 37:5:10). The input resolution and the total number of epochs are set as 32×320×320 and 300, with a batch size of 4. The initial learning rate is set to 0.001 and the Adam algorithm is selected as the optimizer. The optimal hyperparameters for testing are specified through the validation set. The Dice coefficient (Dice) and Intersection over Union (IoU) are adopted as the evaluation indicators. All results are obtained by averaging five different random seed runs to make the experimental results more convincing.

\subsection{Comparison with State-of-the-art Methods}
\label{ssec:Comparison}

The Dice coefficient of liver segmentation results based on the 3D U-Net is 0.9517 and includes all tumor regions.

 As shown in Table \ref{Tab1}, our model achieves the best test scores of 0.8496 and 0.6747 for simultaneous segmenting of liver tumors and intrahepatic vessels. Previous approaches have used complex network designs (\eg, residual connection \cite{3dresunet2019} and  multi-head self-attention \cite{uxnet2022}) to introduce huge parameters to achieve less satisfactory results (\ie, about 1-3\% gains). In contrast, as a 3D network structure, the UCA-Net achieves a vast improvement with a minimal number of parameters similar to 2D networks with only a shallower network layer and a smaller number of channels. Our model significantly outperforms the 2D model, with a 6.97-8.70\% improvement in the tumor segmentation task and a 4.85-7.17\% improvement in the vessel segmentation task. In addition, we also get consistent improvements over 3D models.

Our method is significantly better than the 2D model. The main reason is that the encoder features retained by cross-attention can be effectively combined with the decoder features to take full advantage of 3D context information. For the 3D model, the UCA-Net can adaptively focus on the more valuable information of the encoder and better guide the decoding process during the training to improve the model's performance.

\subsection{Visualization of Segmentation Results}
\label{ssec:Visualization}
The segmentation results are visualized in Figure 3. On the whole, the visualization effect of our model is significantly better than other models. As seen from the segmentation results of terminal vessels shown in the orange dashed box, our model does not confuse blood vessels and tumors like 2D networks, and the segmentation is more refined than other 3D models. In addition, the gold dotted box also shows that UCA-Net segmentation of tumor boundary is more accurate from a 2D view. Finally, as shown in the solid red box, it is easy for 3D models to ignore small tumors in multiple tumor data, but our model performance improved dramatically.

\subsection{Ablation Study}
\label{ssec:Ablation}
We perform ablation studies on the proposed UCA-Net, presented in Table 2. The results prove the necessity of the CCA and SCA modules in the UCA-Net. When only the CCA module is used, tumor and blood vessel segmentation results are improved by 2.37\% and 1.40\%, respectively. The cross-attention mechanism can adaptively focus on the valuable information with a higher correlation degree in the decoder, thus improving network training performance. 
The SCA module focuses on the ability to represent the context semantics between slices, which is consistent with the doctors' habit of labeling tumors slice by slice according to the axial position of tumors, thus achieving a considerable improvement of 3.84\%. However, the growth direction of intrahepatic blood vessels is very complex, which conflicts with the network training that only focuses on the relationship between axial sections. Finally, the combination of CCA and SCA modules not only ensures the consistency of features between the encoder and decoder but also significantly avoids the neglect of blood vessel segmentation while improving the performance of tumor segmentation by the network.

\section{CONCLUSION}
\label{sec:CONCLUSIONS}

To further advance the accurate segmentation of liver tumors and intrahepatic vessels, we collect the LTVS dataset containing 16,052 slices from 52 patients. The dataset has a rich diversity and excellent labeling, promoting community scholars' study of peritumoral vessels. Furthermore, the UCA-Net model based on the customized cross-attention mechanism is proposed to consistently improve the simultaneous segmentation task of liver tumors and intrahepatic vessels. Our model represents a new way to effectively connect the features of the encoder and decoder, where the CCA module can make the network focus on more valuable features, and the SCA module can significantly improve the ability of the network to represent 3D semantic information.
The experimental results have verified the UCA-Net's effectiveness and the components' necessity.

% \vfill
% \pagebreak
\vfill\pagebreak

\bibliographystyle{IEEEbib}
\bibliography{refs}

\end{document}